\def\year{2020}\relax
\documentclass[letterpaper]{article} % DO NOT CHANGE THIS
\usepackage{aaai20}  % DO NOT CHANGE THIS
\usepackage{times}  % DO NOT CHANGE THIS
\usepackage{helvet} % DO NOT CHANGE THIS
\usepackage{courier}  % DO NOT CHANGE THIS
\usepackage[hyphens]{url}  % DO NOT CHANGE THIS
\usepackage{graphicx} % DO NOT CHANGE THIS
\usepackage{graphicx}  % DO NOT CHANGE THIS
\usepackage{amsmath,bm}
\usepackage{amsfonts}
\usepackage{multirow}
\usepackage{comment}

\title{Neural Particle Image Velocimetry}
\author{Nikolay Stulov\textsuperscript{\rm 1}\\
\textsuperscript{\rm 1}Skolkovo Institute of Science and Technology\\
Moscow, 121205, Russia, \\
nikolai.stulov@skoltech.ru
\And Michael Chertkov\textsuperscript{\rm 2,1}\\
\textsuperscript{\rm 2}Program in Applied Mathematics, University of Arizona\\
Tucson, AZ 85721, USA\\
chertkov@arizona.edu
}

\usepackage[dvipsnames]{xcolor}
\usepackage{xspace}
\newcommand{\misha}[1]{{\color{Maroon} Misha says: #1\xspace}}

\begin{document}

\maketitle

\begin{abstract}
In the past decades, great progress has been made in the field of optical and particle based measurement techniques for experimental analysis of fluid flows. Particle Image Velocimetry (PIV) technique is widely used to identify flow parameters from time-consecutive snapshots of particles injected into the fluid. The computation is performed as post-processing of the experimental data via proximity measure between particles in frames of reference.

However, the post-processing step becomes problematic as the motility and density of the particles increases, since the data emerges in extreme rates and volumes. Moreover, existing algorithms for PIV either provide sparse estimations of the flow or require large computational time frame preventing from on-line use.

The goal of this manuscript is therefore to develop an accurate on-line algorithm for estimation of fine-grained velocity field from PIV data. As the data constitutes a pair of images, we employ computer vision methods to solve the problem.

In this work we introduce a convolutional neural network adapted to the problem, namely Volumetric Correspondence Network (VCN) which was recently proposed for the end-to-end optical flow estimation in computer vision. The network is thoroughly trained and tested on a dataset containing both synthetic and real flow data. Experimental results are analyzed and compared to that of conventional methods as well as other recently introduced methods based on neural networks. Our analysis indicates that the proposed approach provides improved efficiency also keeping accuracy on par with other state-of-the-art methods in the field. We also verify through a-posteriori tests that our newly constructed VCN schemes are reproducing well physically relevant statistics of velocity and velocity gradients.
\end{abstract}

\section{Introduction}

In the past decades, great progress has been made in the field of optical and particle-based measurement techniques for non-intrusive turbulent flow monitoring. Techniques like Particle Image Velocimetry (PIV) \cite{adrian1984scattering} and Particle Tracking Velocimetry (PTV) \cite{adamczyk19882} are widely used to identify flow parameters from time-consecutive snapshots of particles injected in the flow.

In this work we consider the following PIV experimental setting:
\begin{itemize}
    \item Many small particles not affecting the flow but advected by the flow are injected. 
    \item The flow domain is illuminated with laser rays to highlight the particles.
    \item Successive snapshots of the the continuous optical field, including highlighted particles, are recorded by a high-speed camera.
    \item The recorded data is processed to extract information about the flow.
\end{itemize}
See \cite{raffel2018particle,adrian2011particle} for review of the state-of-the-art in PIV.

Consider the example of a stationary incompressible velocity field, 
${\boldsymbol v}(t,{\boldsymbol r})$, in a finite domain of the two-dimensional space, ${\boldsymbol r}\in {\cal D}\subset \mathbb{R}^2$, such that $\forall {\boldsymbol r}\in {\cal D}: {\boldsymbol \nabla}\cdot {\boldsymbol v}=0$. Assume that multiple particles are injected and advected by the flow. We observe particles in the form of an image, which can be considered as a scalar density field $\rho(t, {\boldsymbol r})$, where $t$ is time. We assume that effect of molecular diffusion on the density field, controlled by the diffusion coefficient, $\kappa$, is significantly weaker than the effect of advection. In this setting the following advection-diffusion equation governs the spatio-temporal dynamics of the density field:
\begin{equation}\label{eq:advection-diffusion}
\partial_t\rho(t, {\boldsymbol r})+({\boldsymbol v}\cdot{\boldsymbol \nabla}_{\boldsymbol r})\rho(t, {\boldsymbol r})=\kappa \Delta \rho(t, {\boldsymbol r}).
\end{equation}
\begin{comment}
\misha{
%I suggest to add diffusion term with small diffusivity coefficient and call the resulting equations advection-diffusion equations. Without diffusion the advection equations are singular. Diffusion should be viewed as physical (actually present in the flows) regularization of the singularities.
We may also want to comment later on (in the path forward) about learning advection and diffusion simultaneously.
}
\end{comment}

The PIV problem becomes: given a pair or sequence of successive images $\rho(t_k, {\boldsymbol r})$, to learn the spatial distribution of ${\boldsymbol v}$. This post-processing step, that is, estimation of ${\boldsymbol v}({\boldsymbol r})$ from a pair or sequence of snapshots, sets up the main challenge in PIV.

Two families of methods to approach the challenge are documented in the literature: the Cross-Correlation (CC) based \cite{goodman2005introduction} and the Optical Flow (OF) based \cite{horn1981determining} methods. In fact the two methodologies are complementary, in the sense that CC-based methods are advantageous in terms of the computational efficiency, but produce a rather coarse velocity, while OF-based result in a much better resolved velocity field, but lack efficiency. See \cite{liu2015comparison} reviewing applications of the families of CC-based and OF-based methods to the PIV setting in fluid mechanics.

%\misha{I suggest to add sub-title "Our contribution" and have the following three paragraphs as itemized bullets.}

%\kolya{I will add subsections.}

\subsection{Our Contributions}

We propose to view  density field as a passive scalar, that is a scalar advected by the velocity but not influencing the velocity field. Our PIV task becomes, given consecutive snapshots of the passive scalar (the density field) to reconstruct the velocity.

Density of particles in the PIV data resolved at the pixel level is discontinuous making velocity reconstruction problematic. We resolve the problem coarsening images with a hybrid of CC- and OF-based methods utilizing Convolutional Neural Networks (CNN).

The essence of this work is in adapting to PIV and experimenting with the most recent and advanced VCN architecture, developed for non-physical stereo matching problem. The network is thoroughly trained and tested on a dataset containing both synthetic and real flow data. Experimental results are analyzed and compared to conventional PIV methods as well as to aforementioned modern PIV methods developed recently. This analysis indicates that the proposed approach delivers gain in efficiency while keeping the accuracy on par with the state-of-the-art in PIV.

The material in the manuscript is organized as follows: state of the art in the field is reviewed in Section \emph{Related Work}, which includes (as Subsections) description of \emph{Conventional PIV methods} and \emph{Deep Learning for PIV}. We discuss \emph{data} source we use in the manuscript for training and validations in the Section with aforementioned name. \emph{Our Methodology} and \emph{Experiments and Results}  are described in the following two major Sections of the manuscript. We conclude with the Section devoted to \emph{Conclusions and Path Forward}. 

\section{State of the Art}
\subsection{Conventional PIV methods}

CC-based methods have been around for more than twenty years \cite{goodman2005introduction} and, as such, are extensively studied. The main gist of the method consists in reconstructing correspondences between parts of the two consecutive images by running computations on pairs of interrogation windows (patches) of the two consecutive images separated by a unit of time. The computation on patches is as simple as sum of product of intensities of the two patches, which is referred to as cross-correlation operation further. For a pair of $d$-channel, size-$k$ patches, $\rho_1(\boldsymbol{x}), \rho_2(\boldsymbol{x} + \boldsymbol{u}) \in \mathbb{R}^{d \times (2k+1) \times (2k+1)}$, centered at $\boldsymbol{x}$ and $\boldsymbol{x} + \boldsymbol{u}$ in the first and second images respectively, the cross-correlation operation is defines as
\begin{equation}\label{eq:cross-corr}
    cc(\boldsymbol{u}, \boldsymbol{x}) \doteq \sum_{\boldsymbol{o} \in[-k, k]^2}\left\langle\rho_1\left(\boldsymbol{x} + \boldsymbol{o}\right), \rho_2\left(\boldsymbol{x} + \boldsymbol{u} + \boldsymbol{o}\right)\right\rangle
\end{equation}
where $\langle \cdot, \cdot \rangle$ denotes the channel-wise dot-product.

For each patch of the first image, all patches in the second image are considered, each located at unique displacement $\boldsymbol{u}$. Through computing cross-correlations between each pair, an array of $cc(\boldsymbol{u}, \boldsymbol{x})$ values, called cost volume $C(\boldsymbol{u}, \boldsymbol{x})$, is obtained. Then, an optimal displacement $\boldsymbol{u}$ for location $\boldsymbol{x}$ is extracted by locating the argmaximum of cost volume $C(\boldsymbol{u}, \boldsymbol{x})$:
\begin{equation}\label{eq:cc-opt}
    \boldsymbol{u}_{\boldsymbol{x}}^\ast = \arg\min_{\boldsymbol{u}} cc(\boldsymbol{u}, \boldsymbol{x})
\end{equation}
This operation is repeated for all $\boldsymbol{x}$ associated with locations of patches in the first image. The collected displacements per unit time $\boldsymbol{u}_{\boldsymbol{x}}^\ast$ constitute the predicted velocity at the position $\boldsymbol{x}$. To aid the method, images are warped towards each other according to current flow predictor.

An exemplary CC method is a window deformation iterative multi-grid (WIDIM) method \cite{scarano2001iterative}, which has shown good performance in the International PIV Challenges \cite{stanislas2003main,stanislas2005main,stanislas2008main}.

The optical flow (OF) method, developed by Horn and Schunck in \cite{horn1981determining}, relies on solving a computationally more demanding but also more accurate global optimization 
\begin{equation}\label{eq:horn-schunck}
    \min_{\boldsymbol u} \int_{\cal D}\left\{\|\partial_{t} \rho+({\boldsymbol u}\cdot \boldsymbol{\nabla}) \rho \|_2^{2}+\lambda\|\boldsymbol{\nabla} \cdot \boldsymbol{u}\|_2^{2}\right\} \rm{d}{\boldsymbol r},
\end{equation}
where the objective is the integrated $\ell_2$ norm of the mismatch between the left hand side and the right hand side in the brightness change constraint equation (BCCE), which is essentially the advection-diffusion equation (\ref{eq:advection-diffusion}) with the diffusion coefficient, $\kappa$, replaced by the regularization coefficient, $\lambda$.
\begin{comment}
:
\begin{equation}\label{eq:bcce}
    \partial_{t} \rho+(\boldsymbol{v} \cdot \boldsymbol{\nabla}) \rho = 0,
\end{equation}
which is essentially the advection equation Eq.~\ref{eq:advection-diffusion}. \misha{The regularization in Eq.~(\ref{eq:horn-schunck}) results in the diffusion term in the advection-diffusion equation.}
\end{comment}

If BCCE is satisfied, then under mild assumptions imposed on the sequence of images the functional in Eq.~\ref{eq:horn-schunck} is strictly convex and thus has a unique global optimal point.

To avoid aliasing and improve robustness to the assumption violations, the first step becomes to compute the velocity field $\boldsymbol{d}$ at a coarse level by using only low spatial frequency components. The second step is to compensate the reconstructed motion by warping the images towards each other to obtain $\hat{\rho}$. Then, the higher spatial frequencies are used to estimate a correction optical flow on the warped sequence from Eq.~\ref{eq:hs-refinement}, which results in a refined optical flow estimate. This process should be repeated sequentially for finer spatial scales until the original resolution is reached.
\begin{equation}\label{eq:hs-refinement}
    \min_{\boldsymbol{u}} \int_{\Omega}\left\{\|\partial_{t} \hat{\rho}+(\boldsymbol{u} \cdot \boldsymbol{\nabla}) \hat{\rho} \|_2^{2}+\lambda\|\boldsymbol{\nabla} \cdot (\boldsymbol{u} + \boldsymbol{d})\|_2^{2}\right\} \mathrm{d} \boldsymbol{x}
\end{equation}

The original Horn and Schunck (HS) formulation has been extended and optimized in \cite{ruhnau2005variational}. See also \cite{heitz2010variational} for a comprehensive discussion of the OF methods' state-of-the-art.

Notice that both OF and CC methods were introduced originally for the problem in the field of computer vision for fictitious velocity flow connecting two consecutive images in a movie-type data stream. This problem is named optical flow or stereo matching, and one can view PIV, aimed at reconstructing actual velocity from consecutive images of particles injected in a fluid flow, as a special case of it. The specialization of the PIV problem is due to special constraints related to physical conditions of the actual fluid mechanics experiment. For example, the velocity field may possess a special type/degree of compressibility, or it may show a special multi-scale structure, etc.

\subsection{Deep Learning for PIV}

Notice that neither basic OF nor basic CC provide a satisfactory balance of efficiency and accuracy for stereo matching. It was reported recently, that a novel Deep Learning based extension of the methods is capable to fill this niche. The core idea behind these methods (expressed casually here) consists in incorporating a cross-correlation module (see Eq.~\ref{eq:cross-corr}) and then refining its coarse output through interpolation and regularization modules inspired by the one of the OF-based methods and subject to adjustment (learning). It was made possible by the release of open-source labeled stereo motion estimation datasets \cite{Menze2015CVPR,scharstein2014high,sintel2015}.

In \cite{dosovitskiy2015flownet}, the authors propose two CNN architectures, FlowNetC and FlowNetS. \cite{ilg2017flownet} improves on the result using a stacked architecture called FlowNet2. A different direction was taken in \cite{hui2018liteflownet}, proposing a lightweight yet powerful architecture named LiteFlowNet. Finally, the most recent result, namely Volumetric Correspondence Network (VCN) \cite{yang2019volumetric}, sets a new state-of-the-art in the field by handling the cross-correlation operation result through volumetric 4D convolutions.

Deep Learning methodology was first time applied to the PIV problem in \cite{rabault2017performing}. The next milestone in PIV was achieved in \cite{lee2017piv}, which combined cross-correlation techniques with the power of Deep  Learning. Some of the above-mentioned methods were also successfully adapted to the PIV problem. For example, the authors of \cite{cai2019dense} built a PIV estimator based on FlowNetS architecture, and the authors of \cite{cai2019particle} developed and extended it.

To recap, CNN-based methods provide a dense (per-pixel) motion field similar to OF-based methods at the efficiency of CC-based methods (10-100 ms per image pair). Even though training the model is time consuming, once the network is trained, it can be used for real-time estimation. This feature is very much on demand in fluid mechanics, as it allows real-time (online) monitoring and even active flow control instead of storing and post-processing currently dominating realm of the PIV experiments.

\section{Data}

In this work, we choose to experiment largely with synthetic and experimental data available in open, as well as with the synthetic data we generate ourselves.

\begin{figure}
    \centering
    \includegraphics[width=\linewidth]{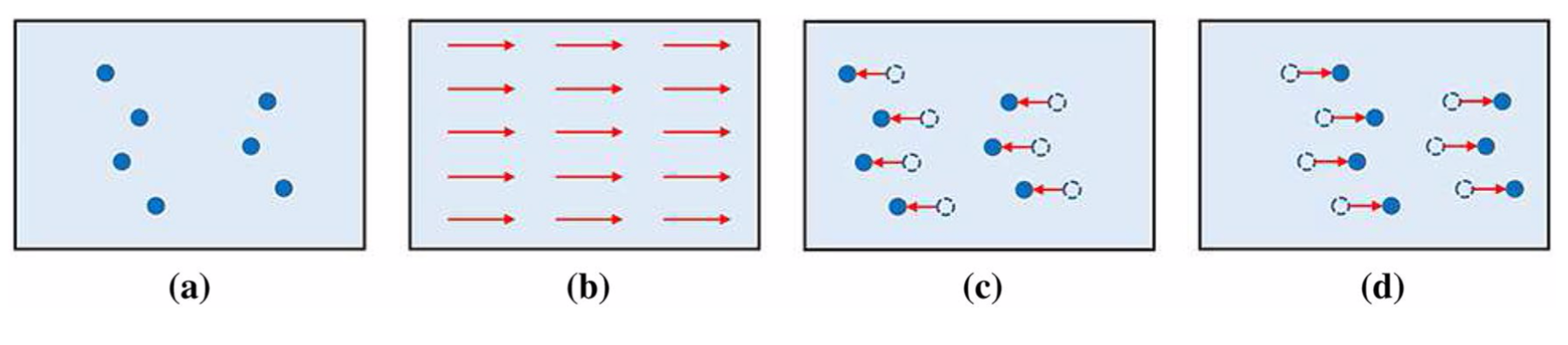}
    \caption{Particle generation procedure from \cite{cai2019dense}. First, particle positions are sampled \textbf{(a)} and the velocity field is prepared \textbf{(b)}. Then, particles are advected by the flow in the forward and backward direction to obtain two snapshots.}
    \label{fig:particle-generation}
\end{figure}

First, we utilize synthetic data assembled in \cite{cai2019dense}, which was generated by mimicking a conventional PIV data-collection procedure as follows. The velocity field is generated or obtained from a database. Consecutive images of the density field are generated by seeding particles homogeneously at random and smoothing the particle representation in the image according to the Gaussian kernel \ref{eq:particle}. Parameters of the kernel are selected randomly from pre-defined ranges described in Tab.~\ref{tab:gaussian-parameters}. Then, particles are advected by the velocity both forward and backward in time to generate synthetic images. The overall procedure is illustrated in 
Fig.~\ref{fig:particle-generation}.

\begin{equation}\label{eq:particle}
    I(x, y)=I_{0} \exp \left(\frac{-\left(x-x_{0}\right)^{2}-\left(y-y_{0}\right)^{2}}{(1 / 8) d_{\mathrm{p}}^{2}}\right).
\end{equation}

\begin{table}
    \centering
    \begin{tabular}{ccc}
        Parameter & Range & Unit \\ \hline
        Seeding density $\rho$ & 0.05 - 0.1 & ppp \\
        Particle diameter $d_p$ & 1 - 4 & pixel \\
        Peak intensity $I_0$ & 200 - 255 & grey value \\
        Location $(x_0, y_0)$ & 1 - 256 & pixel
    \end{tabular}
    \caption{Range of parameter in Eq.~\ref{eq:particle}.}
    \label{tab:gaussian-parameters}
\end{table}

\begin{figure}
    \centering
    \includegraphics[width=\linewidth]{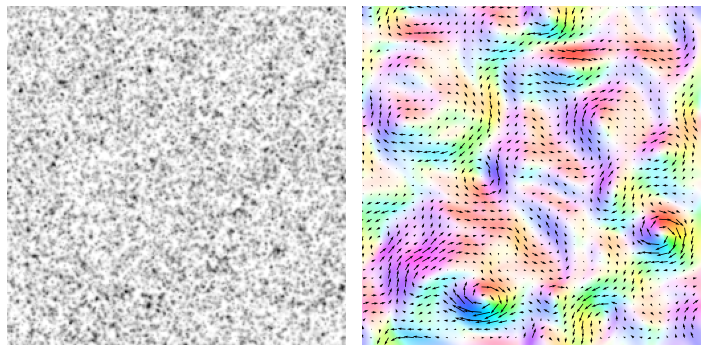}
    \caption{Left: an exemplary density distribution. Right:  respective color-coded velocity field.}
    \label{fig:particle-example}
\end{figure}

An example of an image and the velocity field projecting first snapshot of the density field to the second snapshot are shown in Fig.~\ref{fig:particle-example}.

The velocity field is obtained from five principal sources described in detail in Table~\ref{tab:datasets}. In the table, we denote our simulation of random incompressible flow as {\sc ours}, computational fluid dynamics (CFD) simulation for two flow cases as {\sc uniform}, {\sc back-step} and {\sc cylinder}, 2D turbulent flow motion simulation as {\sc DNS-turbulence}, flow simulation of a surface quasi-geostrophic model as {\sc SQG} and data from the John Hopkins Turbulence Database as {\sc JHTDB}.

We generate our random flow using PhiFlow \cite{holl2020learning} by taking random per-pixel amplitudes and phases in Fourier domain, applying a high-pass filter, smoothing and converting the resulting spectrum to velocity field with inverse Fourier transform. After that, a pressure solver is used to zero out the divergence. This allows us to model an incompressible time-independent multi-scale flow.

The displacements of particles did not exceed 10 pixels in the entire dataset. The use of volumetric convolutions gives premise for adaptation to real data from this limited displacement range.

The overall dataset size is around 15 thousand entries of image pairs and corresponding flows. We use every fifth image pair for testing and the rest for training.

\begin{table*}[ht!]
    \centering
    \begin{tabular}{cccc}
        Name & Description & Condition & Size \\ \hline
        Ours & Random incompressible flow & - & ? \\ \hline
        Uniform & Uniform flow & $|dx| \in [0, 5]$ & 1000 \\ \hline
        \multirow{4}{*}{Back-step} & \multirow{4}{*}{Backward stepping flow} & Re = 800 & 600 \\
        & & Re = 1000 & 600 \\
        & & Re = 1200 & 1000 \\
        & & Re = 1500 & 1000 \\ \hline
        \multirow{5}{*}{Cylinder} & \multirow{5}{*}{Flow over a circular cylinder} & Re = 40 & 50 \\
        & & Re = 150 & 500 \\
        & & Re = 200 & 500 \\
        & & Re = 300 & 500 \\
        & & Re = 400 & 500 \\ \hline
        DNS-turbulence & Homogeneous isotropic turbulent flow & - & 2000 \\ \hline
        SQG & SQG sea surface flow & - & 1500 \\ \hline
        JHTDB-channel & Channel flow provided by JHTDB & - & 1600 \\ \hline
        JHTDB-mhd & Forced MHD turbulence provided by JHTDB & - & 800 \\ \hline
        JHTDB-isotropic & Forced isotropic turbulence provided by JHTDB & - & 2000
    \end{tabular}
    \caption{Description of the data extracted from the John Hopkins Turbulence Database \cite{JHTDB}.}
    \label{tab:datasets}
\end{table*}

\section{Our Methodology}

Let us remind that we consider the PIV problem as the problem of finding (inferring) the velocity field from a pair of filtered images by computing a matching between them, analogous to CC-based methods. A standard filter can be used, however a learned convolutional filter is advantageous as providing greater flexibility. A sequential coarse-to-fine framework, equivalent to one of the methods accumulated within the OF-based methods, yields a spatially well-resolved prediction. In order to assist the network in flow refinement, at every level the first feature map is warped towards the second one with the velocity field  obtained from the previous step.

\begin{figure*}
    \centering
    \includegraphics[width=\linewidth]{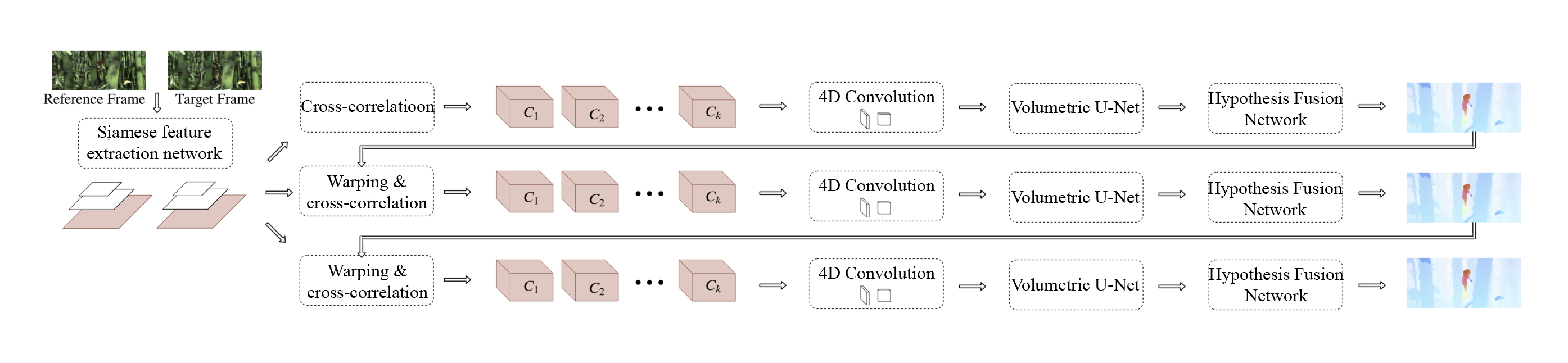}
    \caption{VCN pipeline.}
    \label{fig:vcn-pipeline}
\end{figure*}

VCN improves on the common principal approach described above via several novelties drafted below. We refer the reader to the original VCN paper \cite{yang2019volumetric} for details. The full operational pipeline of VCN is illustrated in Fig.~\ref{fig:vcn-pipeline}.

In the recent literature, the originally 4D cost volume $C(\boldsymbol{u}, \boldsymbol{x})$ is reshaped into 3D by suppressing the two displacement dimensions into one, and processed with 2D convolutions. VCN operates on the original 4D cost volume and processes it with 4D convolutions instead. It allows for volumetric invariance and generalizes well to new unseen displacements. The additional computational burden is resolved with the use of separable 4D convolutions. Application of the 4D filter $K(\boldsymbol{u}, \boldsymbol{x})$ to the cost volume reduces to respective 2D spatial filter $K_S(\boldsymbol{x})$ application and 2D WTA filter $K_{WTA}(\boldsymbol{u})$ application.

Another improvement of VCN is in the cost volume itself. Instead of choosing between the restricted form of the traditional cost volume \cite{chang2018pyramid} and computationally prohibitive feature form of the cost volume \cite{kendall2017end}, VCN introduces a compromise multi-channel cost volume. Thus, instead of a convolution over features in Eq.~\ref{eq:cross-corr}, a feature-wise cosine similarity is used. The final prediction is computed by filtering the hypotheses with channel-wise convolutional modules and weighting results with the soft-max distribution.

We denote the original VCN model, trained for PIV problem, as PIV-VCN. We discover, through extensive experimentation, that the PIV-VCN provides fast and accurate predictions outperforming the other models error-wise. However, visually, the predicted velocity field still has plenty of room for improvement, especially at the finer scales and mostly localized in the areas of large vorticity.

We conjecture that the caveat in the quality of the velocity field reconstruction is due to the fact that VCN seeks for an outcome resolved only at a quarter of the original image. We verify the hypothesis by adding another iteration of refinement and regularization, so that the prediction of the model is half the resolution of the original image. The refined model is coined PIV-VCN-en in the following.

\section{Experiments and Results}
\subsection{Experimental setup}

Similar to PWC-Net and LiteFlowNet \cite{hui2018liteflownet}, VCN uses pyramidal feature extraction through a modified PSPNet \cite{zhao2017pyramid} with a total of 6 levels. Each level is equipped with a coarse-to-fine framework and feature warping, shown in Fig.~\ref{fig:vcn-pipeline}. In PIV-VCN-en, we add additional 7th level at the end. The rest of the hyperparameters for the layers (strides, kernel sizes, hidden layer sizes, number of hypothesis) are set according to the original VCN paper.

Unlike LiteFlowNet, the model is trained simultaneously, using learning rate scheduling with restarts.

As mentioned before, we use 80\% of the dataset for training and 20\% for testing. Moreover, random scale, rotation and translation augmentations are applied to boost the network performance. We found, however, that scaling augmentation prevents the network from learning sub-pixel fine effects of the flow. Thus, we gradually decrease the power of scaling augmentation throughout training.

The network is built and trained in Pytorch. The loss function is a sum of the $\ell_2$ norm contributions correspondent to different scales, each evaluating mismatch  between respective ground truth and predicted flows.

\subsection{Quality }

We utilize  Root Mean Squared Error (RMSE) metric:
\begin{equation}\label{eq:rmse}
    RMSE(x, y) = \sqrt{\frac{1}{N} \sum_{i=1}^N (x_i - y_i)^2},
\end{equation}
to verify quality of the predictions. We also use Squared Error  or $\ell_2$ metric to output pixel-wise error in the Figures presenting our results.

\subsection{Results}
\subsubsection{Qualitative assessment}

\begin{figure}
    \centering
    \includegraphics[width=0.95\linewidth]{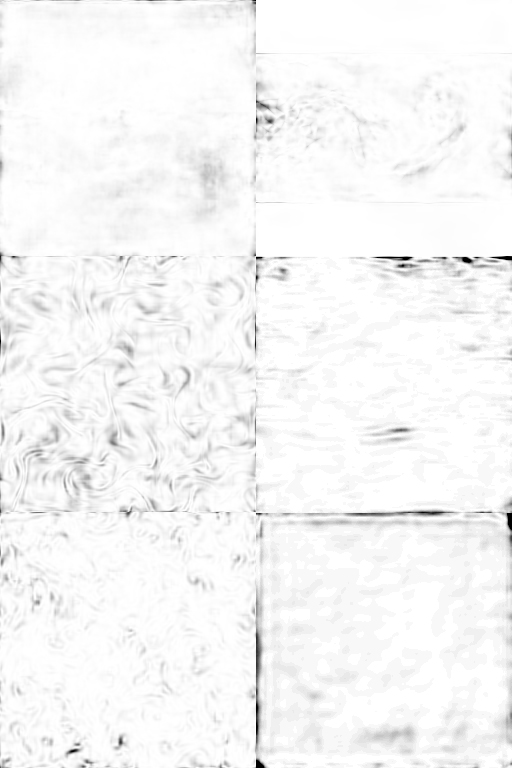}
    \caption{Errors of PIV-VCN model. Left to right, top to bottom: back-step, cylinder, DNS turbulence, JHTDB-channel, SQG, uniform. Darker pixels indicate larger error. For illustration purposes, contrast was enhanced.}
    \label{fig:vcn-errors}
\end{figure}

We show results of the PIV-VCN model error analysis by plotting the squared error of the  reconstructed images in Fig.~\ref{fig:vcn-errors}.

For PIV-VCN model, minor differences between the predicted velocity field and the respective ground truth are visible, especially in the areas of vortexes. However  and by enlarge, the overall quality of estimation is good. Moreover, in the case of the IV-VCN-en model the predicted flow matches the ground truth even better (almost perfectly).  

\subsubsection{Quantitative assessment}

We perform evaluation of the VCN model on the test data and compare the results to other methods in the Table~\ref{tab:vcn-numbers}. From these estimations, we conclude that PIV-VCN model is on-par or better than competitors in most cases, with the only notable exception of the Uniform flow. We conjecture that the exception is due to the fact that errors in the coarse flow are magnified by the interpolation procedure. The conjecture is also supported by the improvement seen in PIV-VCN-en. We conclude that PIV-VCN model is best suited for capturing complex flows.

\begin{table*}
    \centering
    \begin{tabular}{ccccccc}
        Model & Uniform & Back-step & Cylinder & Channel & DNS-turbulence & SQG \\ \hline
        WIDIM & \textbf{0.017} & 0.034 & 0.083 & 0.084 & 0.304 & 0.457 \\ \hline
        HS OF & 0.022 & 0.045 & 0.070 & 0.069 & 0.135 & 0.156 \\ \hline
        PIV-DCNN & 0.018 & 0.049 & 0.100 & 0.117 & 0.334 & 0.479 \\
        PIV-FlowNetS & 0.126 & 0.139 & 0.194 & 0.237 & 0.525 & 0.525 \\
        PIV-FlowNetS-en & 0.059 & 0.072 & 0.115 & 0.155 & 0.282 & 0.294 \\
        PIV-LiteFlowNet & 0.054 & 0.056 & 0.083 & 0.104 & 0.196 & 0.202 \\
        PIV-LiteFlowNet-en & 0.026 & 0.033 & 0.049 & 0.075 & 0.122 & 0.126 \\
        PIV-VCN & 0.265 & 0.024 & 0.025 & 0.085 & 0.099 & 0.103 \\
        PIV-VCN-en & 0.036 & \textbf{0.014} & \textbf{0.017} & \textbf{0.039} & \textbf{0.060} & \textbf{0.067}
    \end{tabular}
    \caption{RMSE in pixels on PIV dataset. %The sections separate different approaches. 
    \textbf{Bold} font highlights the best results.}
    \label{tab:vcn-numbers}
\end{table*}

Another important feature of the PIV algorithm is the computational efficiency of the inference step, expressed in the time required for inference shown in Table~\ref{tab:timing}. We observe that PIV-VCN models are among the top-performing models in terms of efficiency and are capable of online computations. Although PIV-VCN-en model is twice as slow as PIV-VCN, it should be noted that it is much more accurate. Let us also note that the speedup of around 20ms for VCN model is reported in \cite{yang2019volumetric} by implementing the cross-correlation operation in CUDA, which we did not yet implement in our experiments.% \kolya{I should try to use CUDA this time.}

\begin{table*}
    \centering
    \begin{tabular}{ccccccc}
        & VCN-en & VCN & LiteFlowNet-en & FlowNetS-en & HS OF & WIDIM \\ \hline
        Device & GPU & GPU & GPU & GPU & CPU & CPU \\
        Inference time & 0.140 & 0.075 & 0.044 & 0.010 & 1.075 & 0.422
    \end{tabular}
    \caption{Inference time (in seconds) on PIV dataset.}
    \label{tab:timing}
\end{table*}

\subsection{Physics-Informed Diagnostics}

Let us now test how our newly developed ML schemes reproduce physical correlations
characteristic of turbulent flows. We continue to work with the ground truth data for velocity fields, ${\bm v}(t;{\bm r})$, correspondent to various cases of developed turbulence from the Johns Hopkins Turbulence Data Base \cite{li2008public,perlman2007data}, and apply the physics-informed diagnostics of ${\bm v}(t;{\bm r})$ reported in \cite{2018King} to the Machine Learning 
scheme discussed above. These basic physics-informed tests include
\begin{itemize}
    \item Testing the degree of the velocity compressibility, i.e. deviation of $(\nabla\cdot {\bm v})$ from zero (applicable to the cases where the flow we learn is incompressible). This test may also be extended to statistics of the velocity gradient tensor, $m^{\alpha\beta}\doteq \nabla^\alpha v^\beta$ (and not only to its diagonal component controlling incompressibility).
    \item Energy  Spectra, $\langle |{\bm v}_k|^2\rangle$, where  ${\bm v}_k$ is the spatial Fourier transform of ${\bm v}(t,{\bm r})$, over the range of  scales and respective $k$-harmonics. ($\langle\cdots\rangle$ stands for spatial, over snapshot, and temporal, over many snapshots, averaging.)
    \item Testing dependence of the $n$-th order velocity structure function, $\langle |{\bm v}(t,{\bm r}_0+{\bm r})-{\bm v}(t,{\bm r}_0)|^n\rangle$, on  the order $n$ and on 
    the scale $r$. Here, $r$ is assumed scanning through the entire inertial range of 
    turbulence (ranging from the smallest, viscous scale all the way to the largest, energy containing scale).
    \item Testing statistics of the velocity gradient tensor, $m^{\alpha\beta}$, coarse-grained over a scale $r$ from the inertial range of turbulence (i.e. spatially convoluted with a kernel, typically of Gaussian shape, of size $r$). Specifically,  we usually aim to reconstruct statistics of the second, $Q$, and third, $R$, invariants of the coarse-grained velocity tensor.  
\end{itemize}
We report results of the first three tests below. 

These tests of velocity statistics can also be extended to verify statistics of the density field, $\rho(t,{\bm r})$.  We will not present results of the density statistics diagnostics in this paper, however describe the tests below for completeness. 
\begin{itemize}
    \item Density spectra, $\langle |\rho_k|^2\rangle$, where $\rho_k$ is the spatial Fourier transform of $\rho(t,{\bm r})$, over the range of scales and respective $k$-harmonics.
    \item Intermittency of density (typically much stronger pronounced than intermittency of velocity) testing scaling of the density increments, $\langle |\rho(t,{\bm r}'+{\bm r})-\rho(t,{\bm r}')|^n\rangle$, as functions of $n$ and $r$, where $r$ scans scales from the inertial range of turbulence.
    \item Statistics of the density gradient vector, $\nabla\rho$, with a special focus on the tails of the probability distribution. 
\end{itemize}
It is also of interest to test some mixed objects, for example
\begin{itemize}
    \item Mean flux of density fluctuations, $\langle \rho(t,{\bm r}'+{\bm r}) v^\alpha(t,{\bm r}')\nabla^\alpha \rho(t,{\bm r}')\rangle$, as a function of $r$, and respective higher order moments (statistics/intermittency) of the flux. 
\end{itemize}

Notice that the diagnostic is focused on evaluating simultaneous (the same time) statistics of velocity and density. Even richer tests can be designed to analyze  temporal, i.e. different time, correlations in Eulerian (not moving with the flow) and Lagrangian (evolving with the flow) frames. 

The extended diagnostics of turbulence is expected to be useful not only for a-posteriori test of the Machine Learning schemes but also for enforcing expected physical correlations a-priori.  We envision incorporating (in the future work) some of the physics-informed regularizations to the loss function. This approach should help  to minimize deviations of such physically-significant statistical characteristics which show the largest mismatch in the a-posteriori tests.

\begin{figure}[ht!]
    \centering
    \includegraphics[width=0.95\linewidth]{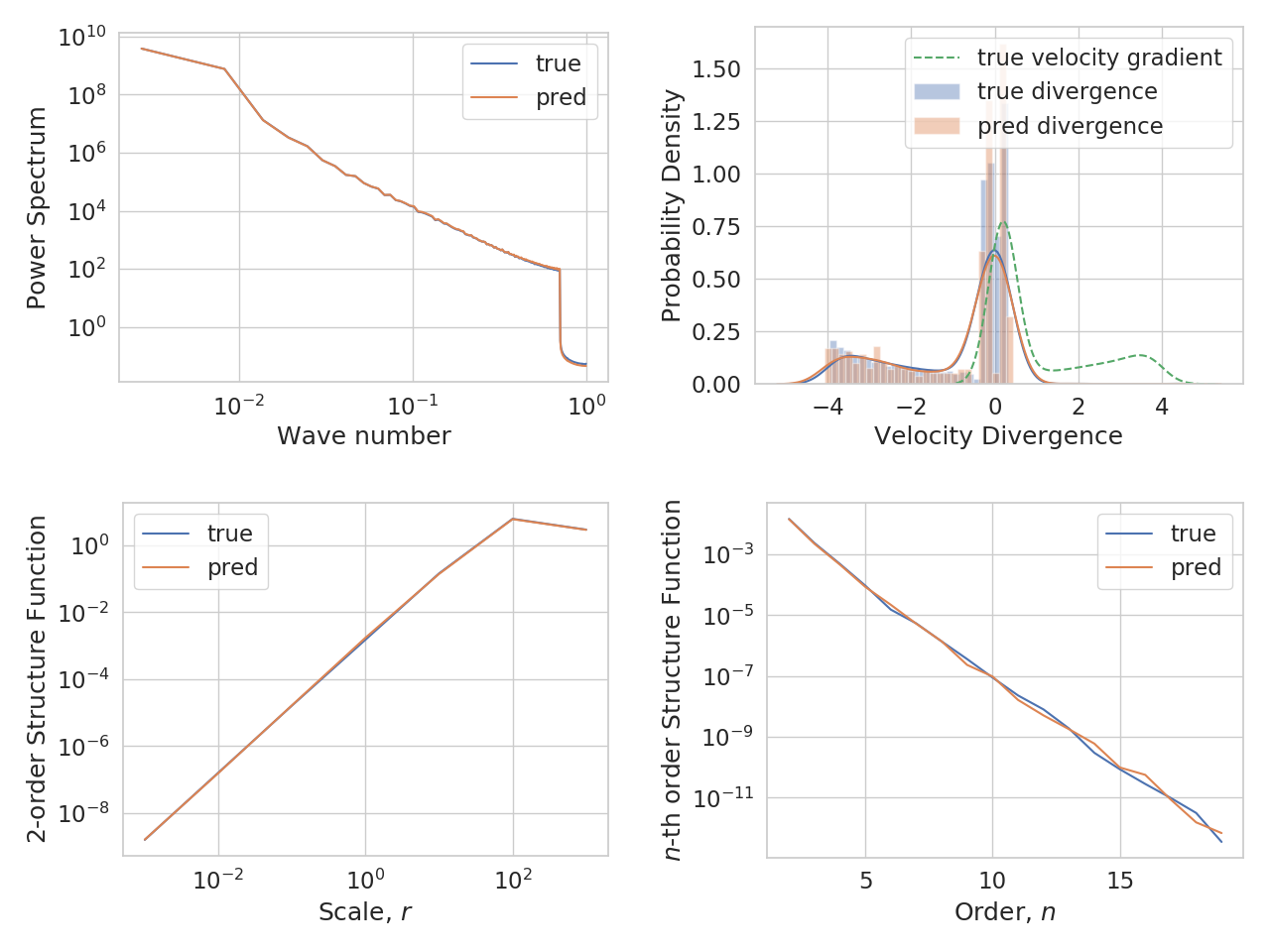}
    \caption{Backward-stepping flow.}
    \label{fig:diagnostics-backstep}
\end{figure}

\begin{figure}[ht!]
    \centering
    \includegraphics[width=0.95\linewidth]{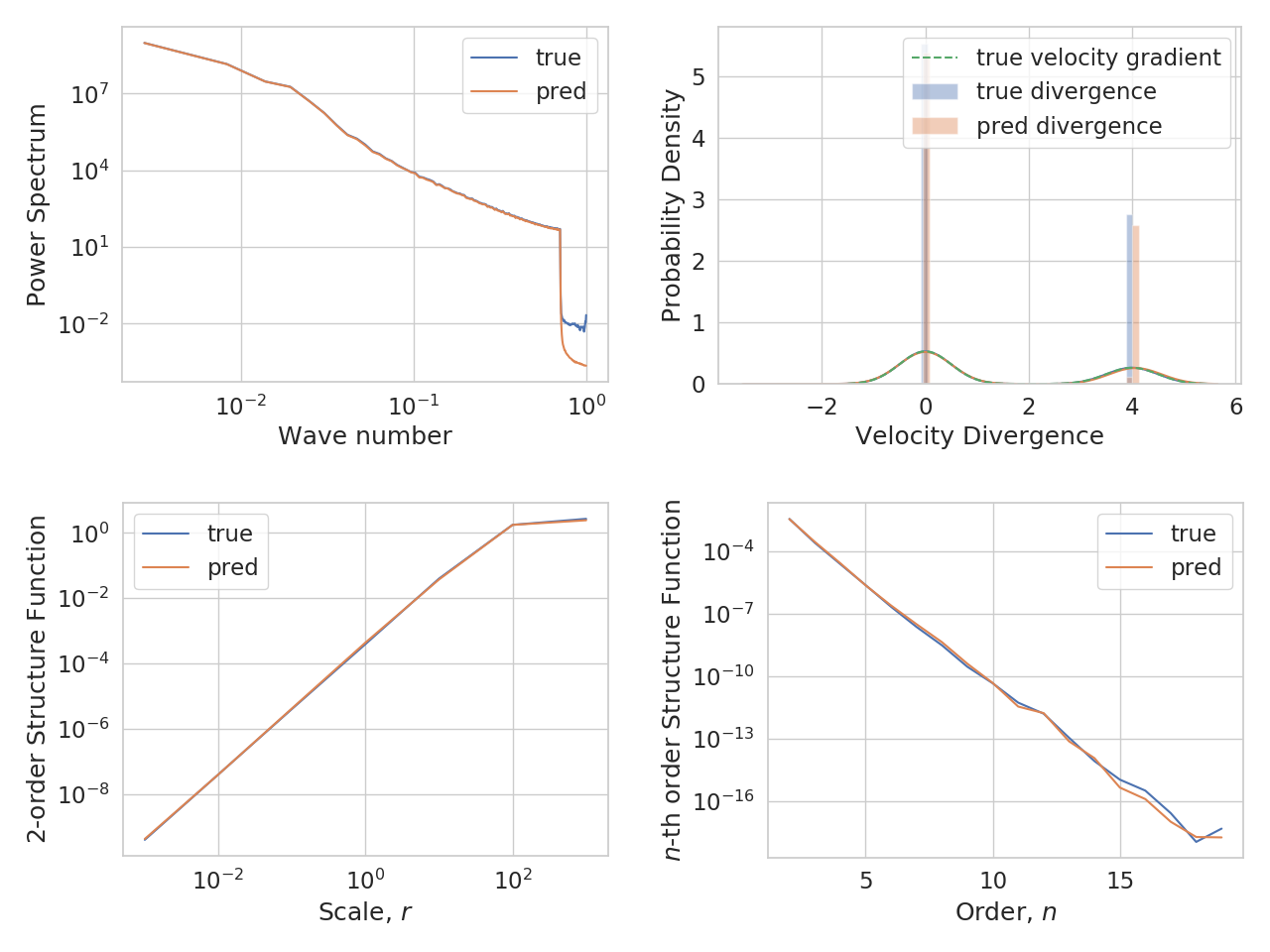}
    \caption{Cylinder flow.}
    \label{fig:diagnostics-cylinder}
\end{figure}

\begin{figure}[ht!]
    \centering
    \includegraphics[width=0.95\linewidth]{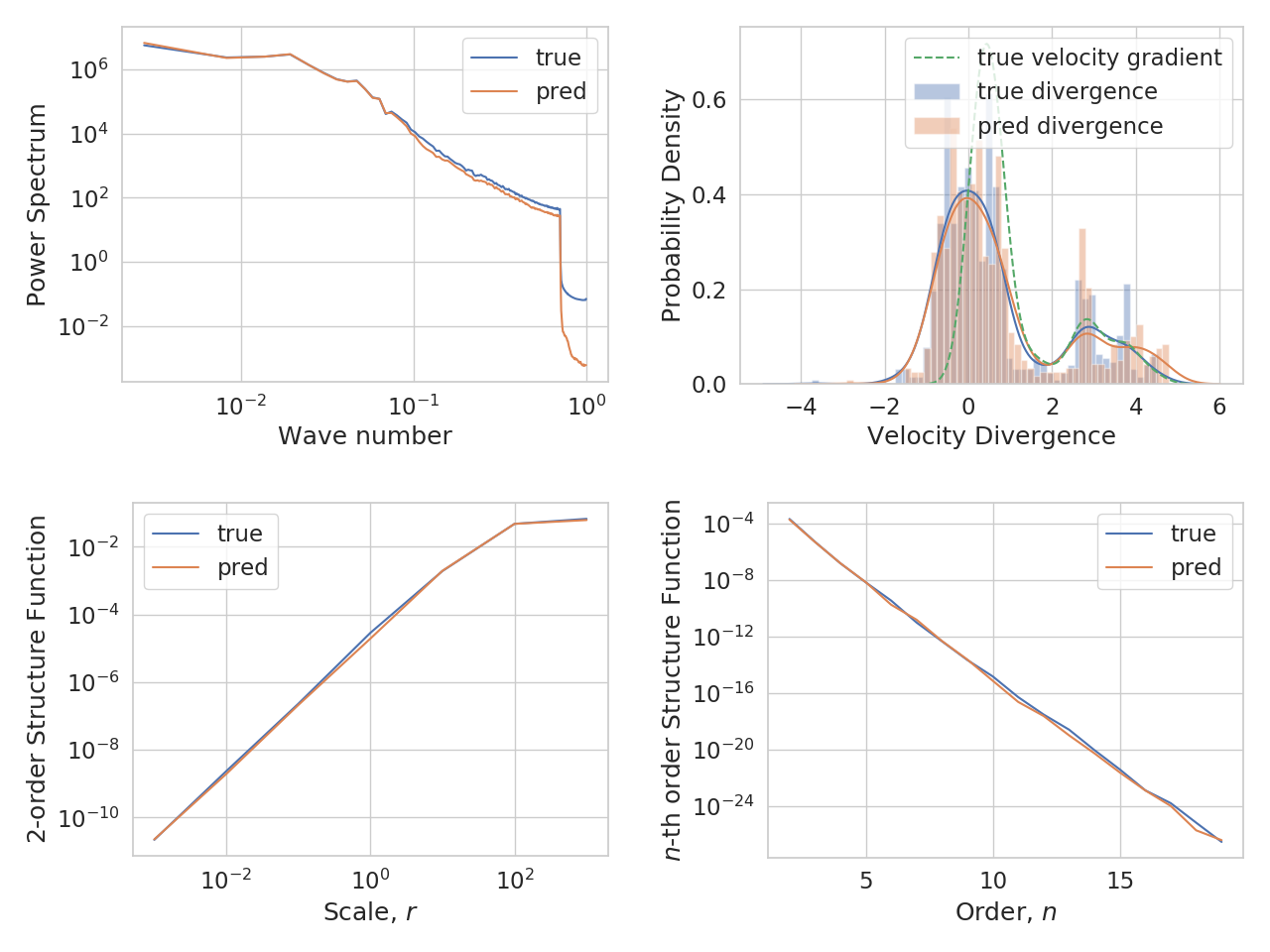}
    \caption{Channel flow.}
    \label{fig:diagnostics-channel}
\end{figure}

\begin{figure}[ht!]
    \centering
    \includegraphics[width=0.95\linewidth]{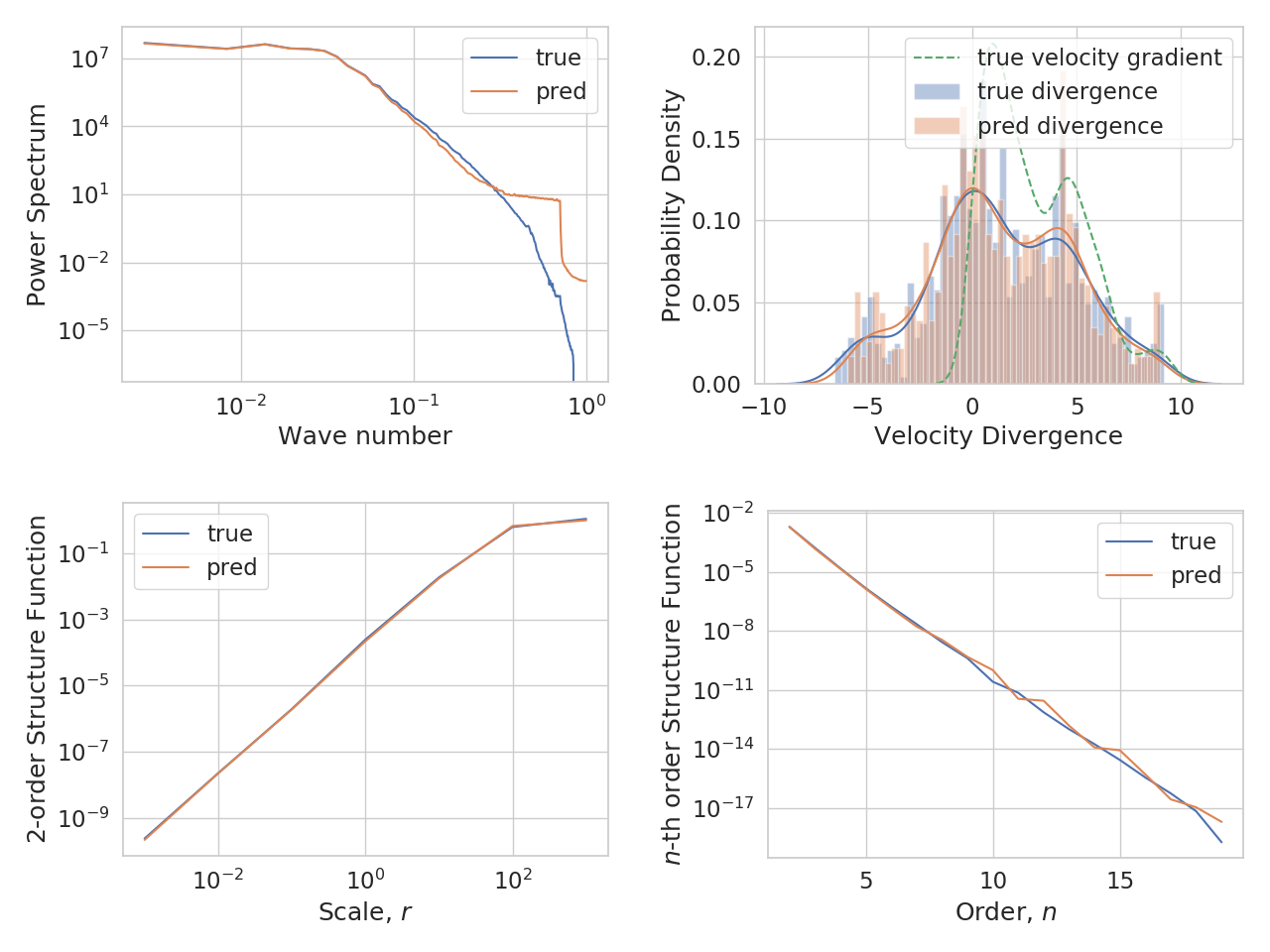}
    \caption{DNS turbulence flow.}
    \label{fig:diagnostics-dns}
\end{figure}

\begin{figure}[ht!]
    \centering
    \includegraphics[width=0.95\linewidth]{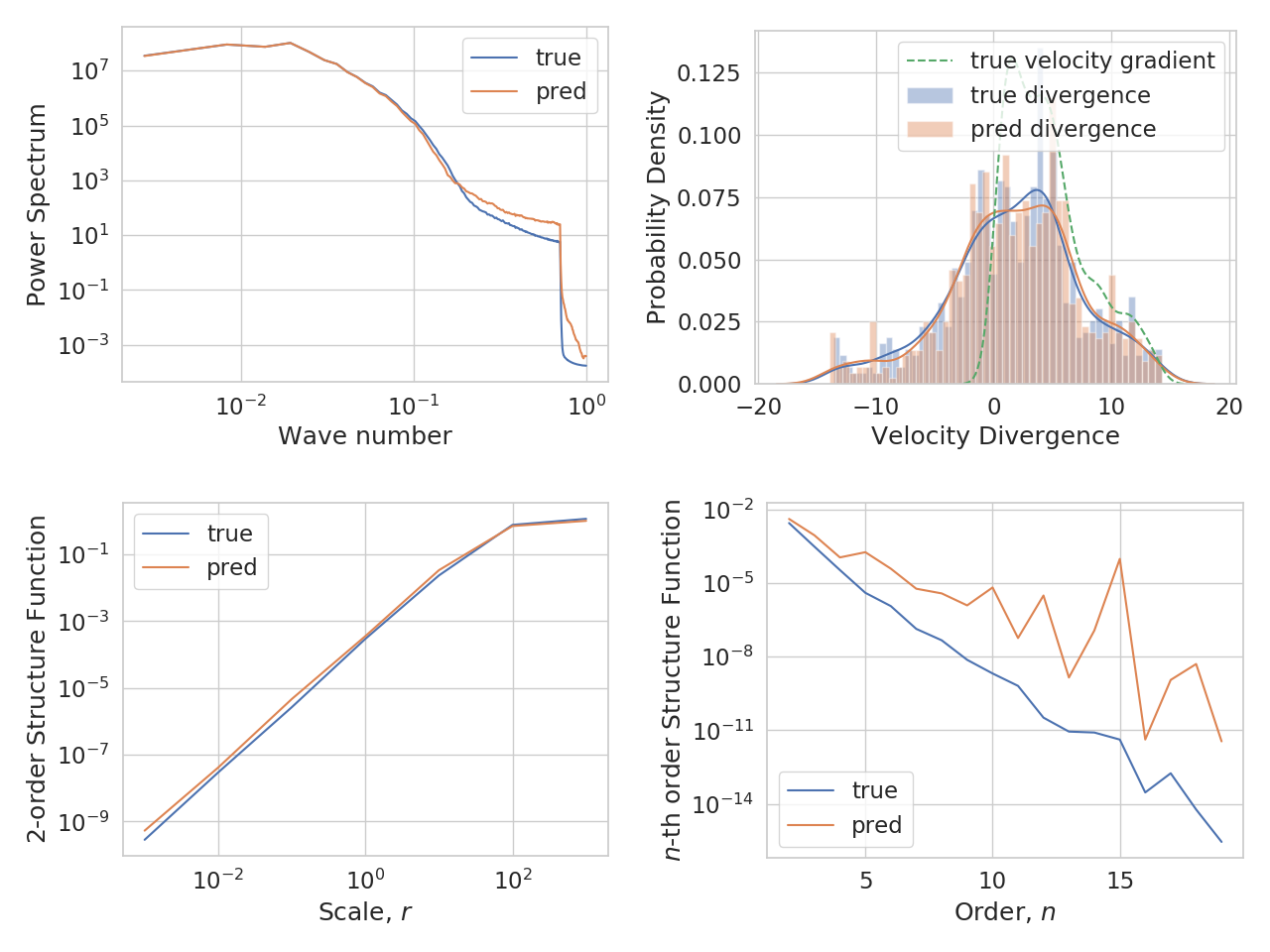}
    \caption{SQG flow.}
    \label{fig:diagnostics-sqg}
\end{figure}

In the rest of the Section we give more details on testing the a-posteriori diagnostics. We present results of the four tests of the simultaneous velocity statistics described above. The tests are applied to five data sets from the JHU turbulence database correspondent to a 
back-step flow (shown in Fig.~\ref{fig:diagnostics-backstep}), a cylinder flow (shown in Fig.~\ref{fig:diagnostics-cylinder}), a channel flow, (shown in Fig.~\ref{fig:diagnostics-channel}), an exemplary homogeneous isotropic (compressible) turbulence flow (shown in Fig.~\ref{fig:diagnostics-dns}) and the surface quasi geostrophic flow (shown in Fig.~\ref{fig:diagnostics-sqg}).

\subsubsection{Power Spectrum Test}

We work with a standard 2D Discrete Fourier Transform of the flow spatial snapshot
%$$F(k_x, k_y) = \sum\limits_{x=1}^{n-1} \sum\limits_{y=1}^{n-1} a(x, y) \exp \left\{-2 \pi i {(x k_x + y  k_y) \over n}\right\},$$
and then average it over the wave number orientation. The resulting power spectrum is shown 
%We obtain the Power Spectrum Profile by averaging the Power Spectrum Density over pixels positions within a certain range of the wave  numbers. We split the continuous space of wave numbers  into bins discrete waves:$$ F(k) = \sum\limits_{m=1}^{m_{\max}} a_m \exp \left\{-2 \pi i k x_m\right\},$$ where $a_m$ are the values of the pixels, $x_m$ are the indices of the pixels, and $k$ is the spatial frequency. The Power Spectrum Profile $F(f)$ 
in the log-log scale against the wave number. The wave number range extends from the largest spatial scales of turbulence (smallest wave numbers), correspondent to the energy containing scales, to the smallest scale (largest wave numbers) correspondent to the viscous (Kolmogorov) scale.  
% \misha{Kolya,  please change it back to the log-log scale.  This is a traditional way of
% the turbulence energy spectrum is shown. See e.g. \url{https://en.wikipedia.org/wiki/Energy_cascade#}.}

\subsubsection{Divergence Test}

We utilize the discrete differences (Sobel operator) to compute the velocity Jacobian $\mathcal{J}(\bm v)$ at each pixel of a snapshot and then use the result to obtain the divergence $(\nabla \cdot {\bm v}) = \mathcal{J}_{xx} + \mathcal{J}_{yy}$. We collect statistics of the divergence averaging over pixel and then compare to the ground truth results with respective outcome of collecting statics over pixels of the trained snapshot. We also compute and show statistics of the velocity gradient, $|\nabla {\bm v}| = \sqrt{\mathcal{J}_{xx}^2 + \mathcal{J}_{yy}^2}$. This is done only for the ground truth data to set up a comparative scale for the divergence of the velocity. Notice,  that of all the examples shown only one,  correspondent to the Cylinder flow, was actually 2d divergence-free. (Majority of the JHU database examples are divergence free,  however in three dimensions.  Therefore,  working with 2d projections we see the flows as 2d-compressible.)

%Our plots show the (empirical) probability distributions of the per-pixel divergences. Probability distribution of the velocity gradient is shown on the same plots for the contrast. 

%It is important to highlight that not all of the flows considered were divergence-free and this test therefore allows us to make assumptions on the possible generalization to arbitrary turbulent flows.

\subsubsection{Structure Function Tests}
We perform two structure function (average velocity increment) tests. First, we consider the second order structure function analyzed as the function of the spatial scale. (Notice,  that the test is actually a $r$-space version of the $k$-space energy spectrum test above.) Then, we set the $r$ scale to three pixels (roughly correspondent to the viscous scale) and analyze how structure function of different orders scale with $n$. In each of the tests we average within a snapshot over the initial position, ${\bm r_0}$, of the pair of points central position and then over orientations of the radius vector connecting the two points. 

%The test against scale is in fact repetitive with respect to the Power Spectrum Profile test, and is given for completeness. In this test, we set the order to 2 and examine the dependency of structure function from scales ranging from smallest (viscous) scales to largest (energy-containing) scales.

%In the test against order, we set the scale to 3 pixel (roughly correspondent to Kolmogorov/viscous scale) and examine the dependency of structure function from order.

%\subsubsection{Results}

%We present our test results in Fig.~\ref{fig:diagnostics-cylinder}-\ref{fig:diagnostics-sqg} for all flows except the uniform flow, which is not turbulent. For each flow, we provide the four tests described above: Power Spectrum Profile test, Divergence test and two Structure Function tests.

Our results show that the ML schemes introduced in this manuscript reproduce statistics of the physically significant characteristics well.  Moreover, performance of the algorithms in terms of passing the physical tests seems better than these discussed in \cite{2018King}. (Notice that this preliminary conclusion will need to be scrutinized in more quantitative future tests.)  We attribute this good performance to the fact that the new ML scheme utilizes sufficient amount of physical information through the scalar - velocity relation.

\section{Conclusion and Path Forward}
\subsection{Conclusion}

In this manuscript, we propose a Deep  Learning estimator for PIV, that is able to extract multi-scale velocity field from consecutive PIV images. Our method is an adaptation of the Volumetric Correspondence Network approach, developed in \cite{yang2019volumetric}, to the PIV setting/data. The resulting model, PIV-VCN-en, is examined in a series of experiments and thoroughly compared to both conventional CC-based and OF-based methods, WIDIM and HS respectively, and to other Deep  Learning-based methods suggested for PIV recently.

The CC-based method lacks estimation accuracy and spatial resolution, while the computation of the OF-based approach is  time-consuming. The proposed approach leverages advantages of both methods. In contrast to WIDIM method, our PIV-VCN-en method provides a much better spatial resolution. Moreover, it is advantageous to the HS method in its computational efficiency. Improvement is also achieved in comparison with the PIV-LiteFlowNet-en approach considered to be the current state-of-the-art in the field.

Finally, we showed that our newly designed Deep Learning PIV estimators, PIV-VCN and PIV-VCN-en, pass an advanced physics-informed turbulence diagnostic tests with flying colors. 

\subsection{Path forward}

Although, as of now, the proposed model does not outperform some state-of-the-art methods in all cases, we intend to continue this work and achieve the goal of winning the competition across the board in the future. However, what is already demonstrated by our analysis is that there is a tremendous opportunity for practical improvements when we take advantage of the very fast pace of developments of the optical flow methods of the computer vision. We envision that when fully developed and validated our methods will allow integrated solutions for on-line, that is real-time, PIV processing. (See, e.g. \cite{06Yu,09Munoz,19Varon} and references there in, for discussions of some early attempts and challenges of using PIV in real-time, in particular for active flow control.) To accomplish this most ambitious goal will require collection of additional experimental data and further developments and tests. %We also plan to develop and document the results in the form of a comprehensive software toolbox.

\bibliographystyle{aaai}
\bibliography{bibliography.bib}

\begin{thebibliography}{}

\bibitem[\protect\citeauthoryear{Adamczyk and Rimai}{1988}]{adamczyk19882}
Adamczyk, A., and Rimai, L.
\newblock 1988.
\newblock 2-dimensional particle tracking velocimetry (ptv): technique and
  image processing algorithms.
\newblock {\em Experiments in fluids} 6(6):373--380.

\bibitem[\protect\citeauthoryear{Adrian, Adrian, and
  Westerweel}{2011}]{adrian2011particle}
Adrian, L.; Adrian, R.~J.; and Westerweel, J.
\newblock 2011.
\newblock {\em Particle image velocimetry}.
\newblock Number~30. Cambridge university press.

\bibitem[\protect\citeauthoryear{Adrian}{1984}]{adrian1984scattering}
Adrian, R.~J.
\newblock 1984.
\newblock Scattering particle characteristics and their effect on pulsed laser
  measurements of fluid flow: speckle velocimetry vs particle image
  velocimetry.
\newblock {\em Applied optics} 23(11):1690--1691.

\bibitem[\protect\citeauthoryear{Butler \bgroup et al\mbox.\egroup
  }{2012}]{sintel2015}
Butler, D.~J.; Wulff, J.; Stanley, G.~B.; and Black, M.~J.
\newblock 2012.
\newblock A naturalistic open source movie for optical flow evaluation.
\newblock In {A. Fitzgibbon et al. (Eds.)}., ed., {\em European Conf. on
  Computer Vision (ECCV)}, Part IV, LNCS 7577,  611--625.
\newblock Springer-Verlag.

\bibitem[\protect\citeauthoryear{Cai \bgroup et al\mbox.\egroup
  }{2019a}]{cai2019particle}
Cai, S.; Liang, J.; Gao, Q.; Xu, C.; and Wei, R.
\newblock 2019a.
\newblock Particle image velocimetry based on a deep learning motion estimator.
\newblock {\em IEEE Transactions on Instrumentation and Measurement}.

\bibitem[\protect\citeauthoryear{Cai \bgroup et al\mbox.\egroup
  }{2019b}]{cai2019dense}
Cai, S.; Zhou, S.; Xu, C.; and Gao, Q.
\newblock 2019b.
\newblock Dense motion estimation of particle images via a convolutional neural
  network.
\newblock {\em Experiments in Fluids} 60(4):73.

\bibitem[\protect\citeauthoryear{Chang and Chen}{2018}]{chang2018pyramid}
Chang, J.-R., and Chen, Y.-S.
\newblock 2018.
\newblock Pyramid stereo matching network.
\newblock In {\em Proceedings of the IEEE Conference on Computer Vision and
  Pattern Recognition},  5410--5418.

\bibitem[\protect\citeauthoryear{Dosovitskiy \bgroup et al\mbox.\egroup
  }{2015}]{dosovitskiy2015flownet}
Dosovitskiy, A.; Fischer, P.; Ilg, E.; Hausser, P.; Hazirbas, C.; Golkov, V.;
  Van Der~Smagt, P.; Cremers, D.; and Brox, T.
\newblock 2015.
\newblock Flownet: Learning optical flow with convolutional networks.
\newblock In {\em Proceedings of the IEEE international conference on computer
  vision},  2758--2766.

\bibitem[\protect\citeauthoryear{Goodman}{2005}]{goodman2005introduction}
Goodman, J.~W.
\newblock 2005.
\newblock {\em Introduction to Fourier optics}.
\newblock Roberts and Company Publishers.

\bibitem[\protect\citeauthoryear{Heitz, M{\'e}min, and
  Schn{\"o}rr}{2010}]{heitz2010variational}
Heitz, D.; M{\'e}min, E.; and Schn{\"o}rr, C.
\newblock 2010.
\newblock Variational fluid flow measurements from image sequences: synopsis
  and perspectives.
\newblock {\em Experiments in fluids} 48(3):369--393.

\bibitem[\protect\citeauthoryear{Holl, Koltun, and
  Thuerey}{2020}]{holl2020learning}
Holl, P.; Koltun, V.; and Thuerey, N.
\newblock 2020.
\newblock Learning to control pdes with differentiable physics.
\newblock {\em arXiv preprint arXiv:2001.07457}.

\bibitem[\protect\citeauthoryear{Horn and Schunck}{1981}]{horn1981determining}
Horn, B.~K., and Schunck, B.~G.
\newblock 1981.
\newblock Determining optical flow.
\newblock In {\em Techniques and Applications of Image Understanding}, volume
  281,  319--331.
\newblock International Society for Optics and Photonics.

\bibitem[\protect\citeauthoryear{Hui, Tang, and
  Change~Loy}{2018}]{hui2018liteflownet}
Hui, T.-W.; Tang, X.; and Change~Loy, C.
\newblock 2018.
\newblock Liteflownet: A lightweight convolutional neural network for optical
  flow estimation.
\newblock In {\em Proceedings of the IEEE conference on computer vision and
  pattern recognition},  8981--8989.

\bibitem[\protect\citeauthoryear{Ilg \bgroup et al\mbox.\egroup
  }{2017}]{ilg2017flownet}
Ilg, E.; Mayer, N.; Saikia, T.; Keuper, M.; Dosovitskiy, A.; and Brox, T.
\newblock 2017.
\newblock Flownet 2.0: Evolution of optical flow estimation with deep networks.
\newblock In {\em Proceedings of the IEEE conference on computer vision and
  pattern recognition},  2462--2470.

\bibitem[\protect\citeauthoryear{{Iriarte Munoz} \bgroup et al\mbox.\egroup
  }{2009}]{09Munoz}
{Iriarte Munoz}, J.~M.; {Dellavale}, D.; {Sonnaillon}, M.~O.; and {Bonetto},
  F.~J.
\newblock 2009.
\newblock Real-time particle image velocimetry based on fpga technology.
\newblock In {\em 2009 5th Southern Conference on Programmable Logic (SPL)},
  147--152.

\bibitem[\protect\citeauthoryear{JHTDB}{}]{JHTDB}
JHTDB.
\newblock Johns {H}opkins {T}urbulence {D}ata {B}ase.
\newblock \url{http://turbulence.pha.jhu.edu/}.

\bibitem[\protect\citeauthoryear{Kendall \bgroup et al\mbox.\egroup
  }{2017}]{kendall2017end}
Kendall, A.; Martirosyan, H.; Dasgupta, S.; Henry, P.; Kennedy, R.; Bachrach,
  A.; and Bry, A.
\newblock 2017.
\newblock End-to-end learning of geometry and context for deep stereo
  regression.
\newblock In {\em Proceedings of the IEEE International Conference on Computer
  Vision},  66--75.

\bibitem[\protect\citeauthoryear{{King} \bgroup et al\mbox.\egroup
  }{2018}]{2018King}
{King}, R.; {Hennigh}, O.; {Mohan}, A.; and {Chertkov}, M.
\newblock 2018.
\newblock {From Deep to Physics-Informed Learning of Turbulence: Diagnostics}.
\newblock {\em Workshop on Modeling and Decision-Making in the Spatiotemporal
  Domain, NeuralIPS 2018; arXiv:1810.07785}.

\bibitem[\protect\citeauthoryear{Lee, Yang, and Yin}{2017}]{lee2017piv}
Lee, Y.; Yang, H.; and Yin, Z.
\newblock 2017.
\newblock Piv-dcnn: cascaded deep convolutional neural networks for particle
  image velocimetry.
\newblock {\em Experiments in Fluids} 58(12):171.

\bibitem[\protect\citeauthoryear{Li \bgroup et al\mbox.\egroup
  }{2008}]{li2008public}
Li, Y.; Perlman, E.; Wan, M.; Yang, Y.; Meneveau, C.; Burns, R.; Chen, S.;
  Szalay, A.; and Eyink, G.
\newblock 2008.
\newblock A public turbulence database cluster and applications to study
  lagrangian evolution of velocity increments in turbulence.
\newblock {\em Journal of Turbulence} (9):N31.

\bibitem[\protect\citeauthoryear{Liu \bgroup et al\mbox.\egroup
  }{2015}]{liu2015comparison}
Liu, T.; Merat, A.; Makhmalbaf, M.; Fajardo, C.; and Merati, P.
\newblock 2015.
\newblock Comparison between optical flow and cross-correlation methods for
  extraction of velocity fields from particle images.
\newblock {\em Experiments in Fluids} 56(8):166.

\bibitem[\protect\citeauthoryear{Menze and Geiger}{2015}]{Menze2015CVPR}
Menze, M., and Geiger, A.
\newblock 2015.
\newblock Object scene flow for autonomous vehicles.
\newblock In {\em Conference on Computer Vision and Pattern Recognition
  (CVPR)}.

\bibitem[\protect\citeauthoryear{Perlman \bgroup et al\mbox.\egroup
  }{2007}]{perlman2007data}
Perlman, E.; Burns, R.; Li, Y.; and Meneveau, C.
\newblock 2007.
\newblock Data exploration of turbulence simulations using a database cluster.
\newblock In {\em Proceedings of the 2007 ACM/IEEE conference on
  Supercomputing},  1--11.

\bibitem[\protect\citeauthoryear{Rabault, Kolaas, and
  Jensen}{2017}]{rabault2017performing}
Rabault, J.; Kolaas, J.; and Jensen, A.
\newblock 2017.
\newblock Performing particle image velocimetry using artificial neural
  networks: a proof-of-concept.
\newblock {\em Measurement Science and Technology} 28(12):125301.

\bibitem[\protect\citeauthoryear{Raffel \bgroup et al\mbox.\egroup
  }{2018}]{raffel2018particle}
Raffel, M.; Willert, C.~E.; Scarano, F.; K{\"a}hler, C.~J.; Wereley, S.~T.; and
  Kompenhans, J.
\newblock 2018.
\newblock {\em Particle image velocimetry: a practical guide}.
\newblock Springer.

\bibitem[\protect\citeauthoryear{Ruhnau \bgroup et al\mbox.\egroup
  }{2005}]{ruhnau2005variational}
Ruhnau, P.; Kohlberger, T.; Schn{\"o}rr, C.; and Nobach, H.
\newblock 2005.
\newblock Variational optical flow estimation for particle image velocimetry.
\newblock {\em Experiments in Fluids} 38(1):21--32.

\bibitem[\protect\citeauthoryear{Scarano}{2001}]{scarano2001iterative}
Scarano, F.
\newblock 2001.
\newblock Iterative image deformation methods in piv.
\newblock {\em Measurement science and technology} 13(1):R1.

\bibitem[\protect\citeauthoryear{Scharstein \bgroup et al\mbox.\egroup
  }{2014}]{scharstein2014high}
Scharstein, D.; Hirschm{\"u}ller, H.; Kitajima, Y.; Krathwohl, G.;
  Ne{\v{s}}i{\'c}, N.; Wang, X.; and Westling, P.
\newblock 2014.
\newblock High-resolution stereo datasets with subpixel-accurate ground truth.
\newblock In {\em German conference on pattern recognition},  31--42.
\newblock Springer.

\bibitem[\protect\citeauthoryear{Stanislas \bgroup et al\mbox.\egroup
  }{2005}]{stanislas2005main}
Stanislas, M.; Okamoto, K.; K{\"a}hler, C.~J.; and Westerweel, J.
\newblock 2005.
\newblock Main results of the second international piv challenge.
\newblock {\em Experiments in fluids} 39(2):170--191.

\bibitem[\protect\citeauthoryear{Stanislas \bgroup et al\mbox.\egroup
  }{2008}]{stanislas2008main}
Stanislas, M.; Okamoto, K.; K{\"a}hler, C.~J.; Westerweel, J.; and Scarano, F.
\newblock 2008.
\newblock Main results of the third international piv challenge.
\newblock {\em Experiments in fluids} 45(1):27--71.

\bibitem[\protect\citeauthoryear{Stanislas, Okamoto, and
  K{\"a}hler}{2003}]{stanislas2003main}
Stanislas, M.; Okamoto, K.; and K{\"a}hler, C.
\newblock 2003.
\newblock Main results of the first international piv challenge.
\newblock {\em Measurement Science and Technology} 14(10):R63.

\bibitem[\protect\citeauthoryear{Varon, Adler, and Eulalie}{2019}]{19Varon}
Varon, E.; Adler, J.; and Eulalie, Y.
\newblock 2019.
\newblock Adaptive control of the dynamics of a fully turbulent bi-modal wake
  using real-time piv.
\newblock {\em Experiments in Fluids} 60:124.

\bibitem[\protect\citeauthoryear{Yang and Ramanan}{2019}]{yang2019volumetric}
Yang, G., and Ramanan, D.
\newblock 2019.
\newblock Volumetric correspondence networks for optical flow.
\newblock In {\em Advances in Neural Information Processing Systems},
  793--803.

\bibitem[\protect\citeauthoryear{Yu \bgroup et al\mbox.\egroup }{2006}]{06Yu}
Yu, H.; Leeser, M.; Tadmor, G.; and Siegel, S.
\newblock 2006.
\newblock Real-time particle image velocimetry for feedback loops using fpga
  implementation.
\newblock {\em Journal of Aerospace Computing, Information, and Communication}
  3(2):52--62.

\bibitem[\protect\citeauthoryear{Zhao \bgroup et al\mbox.\egroup
  }{2017}]{zhao2017pyramid}
Zhao, H.; Shi, J.; Qi, X.; Wang, X.; and Jia, J.
\newblock 2017.
\newblock Pyramid scene parsing network.
\newblock In {\em Proceedings of the IEEE conference on computer vision and
  pattern recognition},  2881--2890.

\end{thebibliography}

\end{document}